# A BAYESIAN FRAMEWORK FOR ESTIMATING VACCINE EFFICACY PER INFECTIOUS CONTACT[1]

By Yang Yang, Peter Gilbert, Ira M. Longini, Jr.
and M. Elizabeth Halloran

*Fred Hutchinson Cancer Research Center and University of Washington*

In vaccine studies for infectious diseases such as human immunodeficiency virus (HIV), the frequency and type of contacts between study participants and infectious sources are among the most informative risk factors, but are often not adequately adjusted for in standard analyses. Such adjustment can improve the assessment of vaccine efficacy as well as the assessment of risk factors. It can be attained by modeling transmission per contact with infectious sources. However, information about contacts that rely on self-reporting by study participants are subject to nontrivial measurement error in many studies. We develop a Bayesian hierarchical model fitted using Markov chain Monte Carlo (MCMC) sampling to estimate the vaccine efficacy controlled for exposure to infection, while adjusting for measurement error in contact-related factors. Our method is used to re-analyze two recent HIV vaccine studies, and the results are compared with the published primary analyses that used standard methods. The proposed method could also be used for other vaccines where contact information is collected, such as human papilloma virus vaccines.

**1. Introduction.** Two randomized multi-center Phase III preventive HIV vaccine trials were conducted to evaluate the efficacy of two versions of AIDSVAX, a recombinant glycoprotein 120 (rgp120) vaccine developed by VaxGen and designed to provide protective immunity by inducing antibody response. One trial (VAX004) was conducted in adults at risk of sexual transmission in North America and the Netherlands, launched in June, 1998, and the other (VAX003) in injecting drug users (IDUs) in Bangkok, Thailand, started in March, 1999. In analyses using Cox proportional hazards models,

Received September 2007; revised June 2008.
[1]Supported by the National Institute of Allergy and Infectious Diseases Grant R01-AI32042.

the vaccine has been shown to be noneffective in Gurwith et al. (2005) for VAX004 and in Pitisuttithum et al. (2006) for VAX003.

A general definition of vaccine efficacy is $VE = 1 - RR$, where RR is the relative risk of infection for a vaccinated subject compared to that for a control subject. Depending on how risk is defined, various VE measures can be derived. The most frequently used measures were classified by Halloran, Struchiner and Longini (1997) into two categories: conditional on exposure to infection and unconditional, that is, whether the measure is controlling for the frequency and type of contacts that lead to transmission. A contact can be defined as one sexual act of a certain type in the context of VAX004 and as one act of sharing a needle for drug injection in VAX003. The VE measure used in Gurwith et al. (2005) and in Pitisuttithum et al. (2006) falls in the unconditional category. It is of public health interest to re-analyze the two vaccine trials using a VE measure conditional on exposure to infection.

For proper inference conditional on exposure to infection, measurement error in exposure factors should be taken into account. For example, the numbers of needle-sharing acts are often under-reported when IDUs are interviewed [Hudgens et al. (2002)]. Thus, methods depending solely on reported exposure information could be inappropriate. To handle the problem of measurement error, many methods have been introduced [Carroll, Ruppert and Stefanski (1995)]. In the nonparametric setting, Fan and Truong (1993) explored the properties of globally consistent nonparametric regression using deconvolution kernels. Cook and Stefanski (1994) and Carroll et al. (1996) developed the simulation extrapolation method that imposes no assumption on the covariates measured with error and uses resampling to detect the trend of measurement error. Richardson and Green (1997) discussed the use of mixture priors for covariates measured with error in the Bayesian framework, and this method was extended to epidemiological studies with a validation set [Richardson et al. (2002)]. In these two vaccine trials, the exposure factors that are subject to measurement error and that are most vital to parameter estimation are the frequencies and the types of contacts.

In this paper we develop a Bayesian framework under the simple assumption of conditional independence [Richardson and Gilks (1993)] for infectious disease incidence data with contact frequency and type recorded for each observation. Using this Bayesian model, we re-analyze the data from the two AIDSVAX trials. Our primary focus is to estimate the transmission probability and vaccine efficacy per infectious contact, while adjusting for measurement error in contact frequency and type. In addition, these studies provide information to address the following questions that are useful for understanding HIV transmission:

- Is VE modified by the baseline behavioral risk profile?

Table 1
*Two randomized multi-center trials conducted for evaluating the efficacy of AIDSVAX, a recombinant glycoprotein 120 HIV-1 vaccine*

|  | **VAX004** | **VAX003** |
|---|---|---|
| Time of trial | 1998–2002 | 1999–2003 |
| Location | North America and The Netherlands | Bangkok, Thailand |
| Type of transmission | Sexual acts | Sharing needles for drug injection |
| Population size | 5403 | 2527 |
|    Male | 5095 (94%) | 2361 (93%) |
|    Female | 308 (6%) | 166 (7%) |
| Randomization ratio |  |  |
|  (vaccine:placebo) | 2:1 | 1:1 |
| Infected/Randomized |  |  |
|   Placebo | 127/1805 | 105/1260 |
|     Male | 123/1704 | 101/1170 |
|     Female | 4/101 | 4/90 |
|   Vaccine | 241/3598 | 106/1267 |
|     Male | 239/3391 | 100/1191 |
|     Female | 2/207 | 6/76 |
| HIV-1 subtypes |  |  |
|   B | 100% | 33 (78%) |
|   E | 0 | 164 (16%) |
|   Untypeable | 0 | 14 (6%) |

- Is the use of condoms in sexual contacts protective?
- Is sharing needles more risky in prison compared to in the general public?
- Is one subtype of HIV more infectious than another subtype via shared needle injection?

The results are compared to those obtained in Gurwith et al. (2005), Pitisuttithum et al. (2006) and Hudgens et al. (2002).

**2. Data description.** Basic characteristics of the two trials are presented in Table 1. The two trials had similar designs except the ratio of vaccine to placebo recipients. Each subject was enrolled free of HIV infection and received seven injections (study vaccine or placebo) at months 0, 1, 6 and every six months thereafter up to month 30. At each immunization visit and the final visit at month 36, antibody assays of blood samples were performed, and exposure factors, adverse events and social harm events for each participant in the past six months were collected. The primary endpoint of the trials was the detection of HIV-1 infection that is defined as both a positive HIV-1 enzyme immunoassay antibody test and the development of at least two new nonvaccine bands on confirmatory HIV immunoblot.

For trial VAX004, in addition to vaccine status, exposure factors were collected in the form of sexual contact frequencies categorized by the behavioral type of the contact (vaginal, oral or anal), gender of the partner, the infection status of partners reported by the subject (HIV-positive, HIV-negative or unknown), and condom use. To reduce the dimension of parameters, we ignore the effects of behavioral type and gender on transmission probabilities by summing the frequencies over the corresponding categories. As the study participants were mostly men that have sex with men (MSM), with females accounting for only 6% of the population and 1.6% of the infections, we are largely assessing transmission via MSM contacts.

For trial VAX003, the exposure factors of interest are the frequency of injections, the fraction of injections using needles shared with other people, the history of injection in jail or prison (incarceration injection), and the vaccine status. Since one of two HIV-1 subtypes (E and B) was found for most infections, it is possible to estimate the transmission probability and vaccine efficacy for each of the two subtypes, given that reasonable estimates of the prevalences of these subtypes among the IDUs in Bangkok, Thailand, are available. Contact information collected in this study is not as detailed as in VAX004. Both the injection frequency and the fraction using shared needles were reported as a few categories instead of numbers. There are four categories for the injection frequency (none, $< 1$/week, $\geq 1$/week but $< 1$/day, and $\geq 1$/day), to which we assign values $10^{-10}$/day, 0.5/week, 4/week and 1/day respectively. There are five categories for the fraction of injections using shared needles (none, occasionally, half of the time, most and always), to which we assign values 0.5%, 15%, 50%, 85% and 99.5% respectively.

## 3. Methods.

### 3.1. *Model structure.*

Following Richardson and Gilks (1993), we specify three submodels for our Bayesian analysis of the measurement error problem: the regression submodel, the measurement error submodel and the prior submodel. In the type of study we are considering, risk factors and infection status are obtained for each subject over consecutive six-month intervals. Let $N$ be the total number of study participants and $T_i$ be the number of intervals of subject $i$, $i = 1, \ldots, N$. We use data collected from month 6 to month 36, excluding month 0 as an adjustment for left truncation. Visits after the first with positive HIV detection are also excluded from analysis. For notational convenience, we identify the $t$th interval of subject $i$ by $(i, t)$.

#### 3.1.1. *The regression submodel.*

Let $p_0$ be the baseline transmission probability per infectious contact. An infectious contact refers to a contact with an infectious source. Let $n_{it}$ be the number of contacts and $\boldsymbol{x}_{itj} = (x_{itj1}, \ldots, x_{itjK})^\tau$ be the vector of $K$ covariates associated with the $j$th contact in interval $(i, t)$, $j = 1, \ldots, n_{it}$. The covariates associated with a contact may include characteristics of the subject (e.g., vaccine status), the partner (e.g., infection status) and the contact itself (e.g., condom use, incarceration, etc.). To associate the transmission probability with covariates, we consider a logit model:

$$p(\boldsymbol{x}_{itj}) = \text{logit}^{-1}(\text{logit}(p_0) + \boldsymbol{x}_{itj}^\tau \boldsymbol{\theta}), \tag{1}$$

where $\boldsymbol{\theta} = (\theta_1, \ldots, \theta_K)^\tau$ is the coefficient vector with the interpretation that $\exp(\theta_k)$ is the increment in odds of transmission per unit increase in $x_{itjk}$ or the odds ratio (OR) for $x_{itjk} = 1$ relative to $x_{itjk} = 0$ if $x_{itjk}$ is binary. Other regression submodels such as the complementary log-log could also be used. Also frequently used is the multiplicative submodel $p(\boldsymbol{x}_{itj}) = p_0 \exp\{\boldsymbol{x}_{itj}^\tau \boldsymbol{\theta}\}$. However, it is sometimes difficult to guarantee $p(\boldsymbol{x}_{itj}) < 1$ when $p_0$ and $\boldsymbol{\theta}$ are simultaneously sampled. In the context of the two AIDSVAX trials, we use $\text{OR}_{vac}$, $\text{OR}_{con}$ and $\text{OR}_{inc}$ to denote the odds ratios of transmission per infectious contact for vaccination, condom use and incarceration, respectively. The probability of escaping infection in interval $(i, t)$ is

$$Q_{it} = \prod_{j=1}^{n_{it}} (1 - p(\boldsymbol{x}_{itj}) \pi(\boldsymbol{x}_{itj})), \tag{2}$$

where $\pi(\boldsymbol{x}_{itj})$ is the prevalence of infectious contacts among all contacts with covariates $\boldsymbol{x}_{itj}$. As $p(\boldsymbol{x}_{itj})$ and $\pi(\boldsymbol{x}_{itj})$ always appear as a product, they are not estimable at the same time, and $\pi(\boldsymbol{x}_{itj})$ is often assumed known and evaluated from either literature or the data.

As mentioned in the introduction, different measures can be used for vaccine efficacy, depending on the definition of relative risks. A natural choice is the VE per infectious contact with the risks being transmission probabilities per infectious contact as given in (1). However, the relative risk obtained from transmission probabilities per infectious contact depends on not only the vaccine status but also other covariates. Such dependency may not exist in different models. For example, if we assume a multiplicative model $p(\boldsymbol{x}_{itj}) = p_0 \exp(\boldsymbol{x}_{itj}^\tau \boldsymbol{\theta})$, the VE per infectious contact will depend solely on the vaccine status. For the logit model, the dependency could also be minimal if $p(\boldsymbol{x}_{itj})$ is small, where we have VE per infectious contact $\approx 1 - \text{OR}_{vac}$. The approximation holds for the contact types we consider here, and thus, we report $1 - \text{OR}_{vac}$ as the VE per infectious contact for the data analysis.

Expressions (1) and (2) provide a general form for the regression submodel. The exact form is specific to each study, depending on the covariates under consideration, and is described below.

*The North America and Netherlands trial (VAX004).* For trial VAX004, we are interested in the effects of vaccine and condom usage. Let $v_i$ indicate the vaccine status (1: yes, 0: no) and $c_{itj}$ indicate the condom use (1: yes, 0: no) for the $j$th sexual contact in interval $(i, t)$. Let $p_0$ be the transmission probability for a sexual contact without a condom between a placebo recipient and an infected partner. We assume the prevalence, $\pi$, of HIV in contacts is identical for all intervals and is known. The escape probability for interval $(i,t)$ is given by

$$(3) \quad Q_{it} = \prod_{j=1}^{n_{it}}(1-p(v_i,c_{itj})\pi) = (1-p(v_i,1)\pi)^{m_{it}}(1-p(v_i,0)\pi)^{n_{it}-m_{it}},$$

where $p(v_i,c_{itj}) = \text{logit}^{-1}(\text{logit}(p_0) + \theta_v v_i + \theta_c c_{itj})$, $\theta_v$ and $\theta_c$ are the effects of the vaccine and condom use, and $m_{it} = \sum_{j=1}^{n_{it}} c_{itj}$, the total number of contacts with a condom. The probability distribution of the final transmission status, $y_{it}$ (1: infection, 0: escape), is then

$$(4) \quad \Pr(y_{it}|n_{it},m_{it},v_i;p_0,\theta_v,\theta_c) = Q_{it}^{1-y_{it}}(1-Q_{it})^{y_{it}}.$$

*The Thai trial (VAX003).* For this trial, we consider vaccine status, incarceration history of the subject and needle-sharing as covariates. Let $p_0$ be the baseline probability of infection by an injection using a needle shared with an HIV-infected person. Let $u_i$ denote whether the subject had incarceration injection (1: yes, 0: no) during the study, and $s_{itj}$ denote whether the injection was using a shared needle (0: yes, 1: no). Also define $\theta_v$, $\theta_u$ and $\theta_s$ as the effects of the covariates, respectively. We assume that injections using nonshared needles were not infectious. That is, $\theta_s = -\infty$, and the regression submodel is built solely on the $m_{it} = \sum_{j=1}^{n_{it}}(1-s_{itj})$ contacts using shared needles. The probability of escaping infection in interval $(i,t)$ is given by

$$(5) \quad Q_{it} = \prod_{j=1}^{n_{it}}(1-p(v_i,u_i,s_{itj})\pi) = (1-p(v_i,u_i,0)\pi)^{m_{it}},$$

where $p(v_i,u_i,s_{itj}) = \text{logit}^{-1}(\text{logit}(p_0) + \theta_v v_i + \theta_u u_i + \theta_s s_{itj})$. The probability distribution of the final transmission status is the same as (4).

As the HIV subtype was determined for most infected subjects, it is possible to estimate the transmission probability and vaccine efficacy for each subtype. Let $p_0^{(e)}$ ($p_0^{(b)}$) be the baseline probability of infection by an injection using a needle shared with somebody infected with HIV of subtype E (B), $\theta_v^{(e)}$ ($\theta_v^{(b)}$) be the vaccine effects against transmission of subtype E (B), and $\pi^{(e)}$ ($\pi^{(b)}$) be the prevalence of people infected with subtype E (B) among the IDU population. The probabilities of escaping infection from injections using needles shared with infected partners of subtype E and subtype B, respectively, are given by

$$Q_{it}^{(e)} = (1 - \text{logit}^{-1}(\text{logit}(p_0^{(e)}) + \theta_v^{(e)}v_i + \theta_u u_i)\pi^{(e)})^{m_{it}}$$

and

$$Q_{it}^{(b)} = (1 - \text{logit}^{-1}(\text{logit}(p_0^{(b)}) + \theta_v^{(b)}v_i + \theta_u u_i)\pi^{(b)})^{m_{it}}.$$

We assume transmission of subtype E is independent of transmission of subtype B. As infection by both subtypes is rare, we assume an infected subject typed as E (B) must have escaped transmission from infectious contacts of subtype B (E). The probability distribution of the final transmission status can be expressed as

(6)
$$\Pr(y_{it}, \text{subtype} | m_{it}, v_i, u_i; p_0, \theta_v, \theta_u)$$
$$= \begin{cases} Q_{it}^{(e)} Q_{it}^{(b)}, & y_{it} = 0, \\ Q_{it}^{(b)}(1 - Q_{it}^{(e)}), & y_{it} = 1, \text{subtype} = \text{E}, \\ Q_{it}^{(e)}(1 - Q_{it}^{(b)}), & y_{it} = 1, \text{subtype} = \text{B}, \\ 1 - Q_{it}^{(e)} Q_{it}^{(b)}, & y_{it} = 1, \text{subtype} = \text{U}, \end{cases}$$

where "U" stands for "Untypeable."

3.1.2. *The measurement error submodel.* We consider two types of exposure information that are measured with error, the total number of contacts, $n_{it}$, and the number of a particular subset of contacts, $m_{it}$. Let $\tilde{n}_{it}$ and $\tilde{m}_{it}$ be the measured values of $n_{it}$ and $m_{it}$, respectively. As data in the form of counts over time periods often arise from a Poisson process, we assume a Poisson distribution for the true number of contacts $n_{it}$ and an over-dispersed Poisson distribution for the measured number $\tilde{n}_{it}$ during a time interval of length $l_{it}$, given the contact rate $\lambda_{it}$. The reason for an over-dispersion structure is that we want some correction for the potentially under- or over-reported number of contacts, for example, the number of sexual contacts in a single interval was reported as thousands by several subjects in trial VAX004. The histograms of reported contact rates in Figure 1(a) for VAX004 and Figure 1(c) for VAX003 suggested either gamma or log-normal distributions. We use the log-normal distribution for illustration, but compare both in the data analyses. Define $\boldsymbol{n}_i = (n_{i1}, \ldots, n_{iT_i})^\tau$, $\boldsymbol{m}_i = (m_{i1}, \ldots, m_{iT_i})^\tau$, $\tilde{\boldsymbol{n}}_{it} = (\tilde{n}_{i1}, \ldots, \tilde{n}_{iT_i})^\tau$, $\tilde{\boldsymbol{m}}_i = (\tilde{m}_{i1}, \ldots, \tilde{m}_{iT_i})^\tau$ and $\boldsymbol{\lambda}_i = (\lambda_{i1}, \ldots, \lambda_{iT_i})^\tau$. Let $\boldsymbol{1}$ and $\boldsymbol{J}$ denote the vector and matrix, respectively, with all elements being 1, and let $\boldsymbol{I}$ denote the identity matrix. The dimensions of $\boldsymbol{1}$, $\boldsymbol{J}$ and $\boldsymbol{I}$ are clear from the context and are thus suppressed. We choose the following measurement error structure for $n_{it}$:

(7)
$$\boldsymbol{\lambda}_i \sim \text{Log-Normal}(\mu \boldsymbol{1}, \sigma^2 (\rho \boldsymbol{J} + (1 - \rho) \boldsymbol{I})),$$
$$n_{it} \sim \text{Poisson}(\lambda_{it} l_{it}),$$
$$\delta_{it} \sim \text{Gamma}(\phi, \lambda_{it} l_{it} / \phi),$$
$$\tilde{n}_{it} \sim \text{Poisson}(\delta_{it}).$$

An exchangeable within-subject correlation structure is assumed for the contact rates, $\boldsymbol{\lambda}_i$, but other correlation structures could be considered. The magnitude of correlation among elements of $\boldsymbol{\lambda}_i$ is measured by $\rho$, $0 \leq \rho \leq 1$, the correlation coefficient for $\log(\boldsymbol{\lambda}_i)$. We assume unbiasness for the measurement error, as $\mathrm{E}(\tilde{n}_{it}|\lambda_{it}) = \lambda_{it} l_{it} = \mathrm{E}(n_{it}|\lambda_{it})$. The over-dispersion is reflected by $\mathrm{VAR}(\tilde{n}_{it}|\lambda_{it}) = \lambda_{it} l_{it}(1 + \lambda_{it} l_{it}/\phi)$ and is generated by adding the layer of $\boldsymbol{\delta}_i = (\delta_{i1}, \ldots, \delta_{iT_i})^\tau$. The degree of over-dispersion decreases as $\phi$ goes to infinity. By our assumption, $n_{it}$ is conditionally independent of $\tilde{n}_{it}$ given the contact rate $\lambda_{it}$. Zero values of $\tilde{n}_{it}$ are allowed for intervals in which infections happened since only $n_{it}$ is required to be nonzero.

Given $n_{it}$ and $\tilde{n}_{it}$, it is natural to choose binomial distributions for both the true number $m_{it}$ and the measured number $\tilde{m}_{it}$ based on a beta-distributed proportion $\xi_{it}$, which is also suggested by the histograms of reported proportions of contacts with condom use in Figure 1(b) for VAX004 and contacts with needle-sharing in Figure 1(d) for VAX003. Define $\Phi(\cdot)$ as the standard

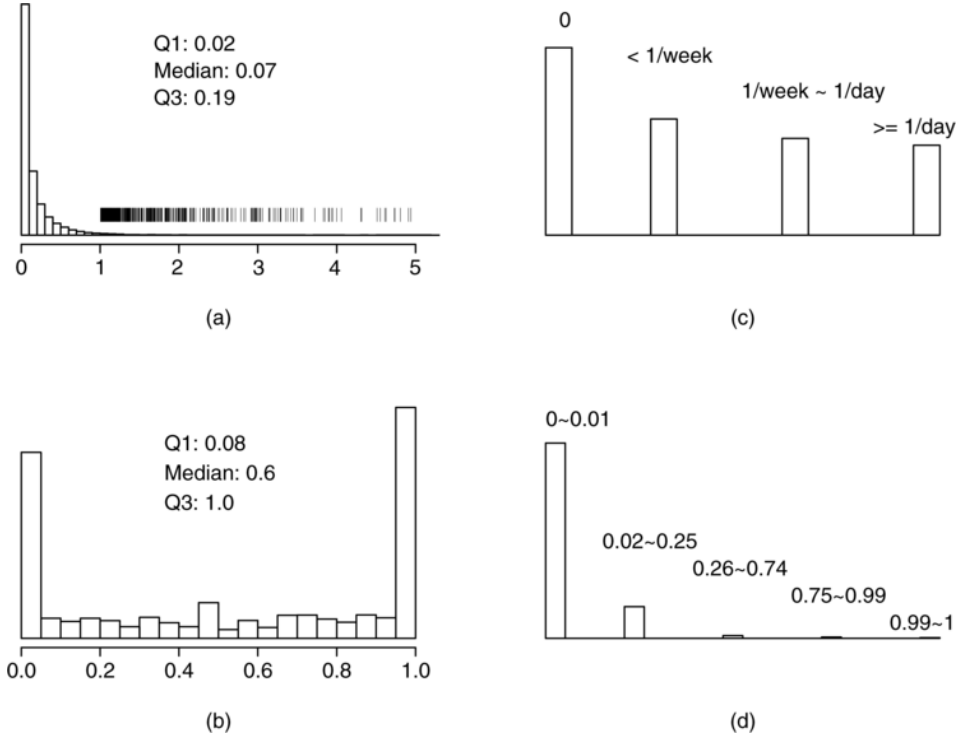

FIG. 1. (a) *Reported sexual contact rates in VAX004. Values larger than 5/day (<0.1%) are truncated in the graph but not in the analysis. The vertical line segments indicate the location of values between 1 and 5.* (b) *Reported proportions of condom use in VAX004.* (c) *Reported injection rates in VAX003.* (d) *Reported proportions of shared needles in VAX003.*

normal cumulative distribution function (CDF) and $\Psi(\cdot|\alpha,\beta)$ as the beta CDF. We have

$$
\begin{aligned}
\xi_{it} &\sim \text{Beta}(\alpha,\beta), \\
m_{it} &\sim \text{Binomial}(n_{it}, \xi_{it}), \\
\tilde{m}_{it} &\sim \text{Binomial}(\tilde{n}_{it}, \xi_{it}), \\
\Phi(\varepsilon_{it}) &= \Psi(\xi_{it}|\alpha,\beta), \\
\boldsymbol{\varepsilon}_i &\sim N(\mathbf{0}, \gamma \boldsymbol{J} + (1-\gamma)\boldsymbol{I}),
\end{aligned}
\tag{8}
$$

where $\boldsymbol{\varepsilon}_i = (\varepsilon_{i1}, \ldots, \varepsilon_{iT_i})^\tau$. We use a standard normal copula to model the within-subject correlation among $\boldsymbol{\xi}_i = (\xi_{i1}, \ldots, \xi_{iT_i})^\tau$, the proportions of contacts in a subcategory (condom use or needle-sharing). This copula is formed by generating a standard normal random vector $\boldsymbol{\varepsilon}_i$ with an exchangeable correlation structure, the correlation coefficient being $\gamma$, and transforming it to a uniform random vector using $\Phi$ on each component. The uniform random vector is then transformed to $\boldsymbol{\xi}_i$ using $\Psi^{-1}$ on each element. The $\boldsymbol{\xi}_i$ generated in this way has marginal CDF $\Psi(\cdot|\alpha,\beta)$ and an exchangeable correlation structure. While the correlation coefficient for $\boldsymbol{\xi}_i$ is not the same as that for $\boldsymbol{\varepsilon}_i$, they share the same rank correlation because the CDFs are monotonic. Note that the log-normal distribution can be viewed as a special case utilizing the standard normal copula. Conditional on $n_{it}$, $\tilde{n}_{it}$ and $\xi_{it}$, $m_{it}$ and $\tilde{m}_{it}$ are independent.

3.1.3. *The prior submodel.* We use the following priors for $p_0$, $\boldsymbol{\theta}$ and hyperparameters:

$$
\begin{aligned}
\mu &\sim 1, \\
\sigma^2 &\sim \frac{1}{\sigma^2}, \\
\rho &\sim \text{Uniform}(0,1), \\
\phi &\sim \left[\ln \Gamma''(\phi) - \frac{1}{\phi}\right]^{1/2}, \\
(\alpha,\beta) &\sim [\ln \Gamma''(\alpha) \ln \Gamma''(\beta) - \ln \Gamma''(\alpha+\beta)(\ln \Gamma''(\alpha) + \ln \Gamma''(\beta))]^{1/2}, \\
\gamma &\sim \text{Uniform}(0,1), \\
\theta_k &\sim \text{Normal}(0, d_k^2), \qquad k = 1, \ldots, K, \\
p_0 &\sim \text{Uniform}(a_p, b_p),
\end{aligned}
\tag{9}
$$

where $\{d_k : k = 1, \ldots, K\}$, $a_p$ and $b_p$ are assumed known, and $\ln \Gamma''(\cdot)$ is the trigamma function. Jeffreys' noninformative priors are used for $\mu$, $\sigma^2$, $\phi$ and $(\alpha,\beta)$.

Our choice of a relatively wide range $(a_p, b_p)$ is guided by the maximum likelihood estimate (MLE) of $p_0$ obtained solely from the regression submodel. To use this simple likelihood method, we assume $n_{it} = \tilde{n}_{it}$, and $m_{it}$ is estimated by $n_{it} \times \sum_{i,t} \tilde{m}_{it} / \sum_{i,t} \tilde{n}_{it}$ for VAX003. The same assumption of a common proportion of shared needles was employed in Hudgens et al. (2002). However, one will not be able to differentiate the condom effect with a common proportion of condom use, and thus, we assume $m_{it} = \tilde{m}_{it}$ additionally to obtain the MLE of $p_0$ for VAX004.

A normal prior $N(0, d_k^2)$ is reasonable for covariate effects because we let the data drive the 95% credible sets away from the null value if strong effects exist. The values of $\{d_k : k = 1, \ldots, K\}$ are set relatively large, for example, 2, to provide a wide domain for the odds ratios.

3.2. *Posterior distributions.* Bayesian inferences are based on posterior distributions of all unknown parameters and latent variables given the data and known parameters, which are derived from the prior and conditional distributions stated in the previous section. Let $\boldsymbol{y} = (\boldsymbol{y}_1^\tau, \ldots, \boldsymbol{y}_N^\tau)^\tau$ be the vector of observed infection status, where $\boldsymbol{y}_i = (y_{i1}, \ldots, y_{iT_i})^\tau$, and let $\boldsymbol{x} = (\boldsymbol{x}_1^\tau, \ldots, \boldsymbol{x}_N^\tau)^\tau$, where $\boldsymbol{x}_i = (\boldsymbol{x}_{i1}^\tau, \ldots, \boldsymbol{x}_{iT_i}^\tau)^\tau$ and $\boldsymbol{x}_{it} = (\boldsymbol{x}_{it1}, \ldots, \boldsymbol{x}_{itn_{it}})^\tau$, be the observed covariate matrix for all intervals. Similarly, define $\boldsymbol{n}$, $\tilde{\boldsymbol{n}}$, $\boldsymbol{m}$, $\tilde{\boldsymbol{m}}$, $\boldsymbol{\lambda}$, $\boldsymbol{\delta}$, $\boldsymbol{\xi}$ and $\boldsymbol{\varepsilon}$ as the vectors of $n_{it}, \tilde{n}_{it}, m_{it}, \tilde{m}_{it}, \lambda_{it}, \delta_{it}, \xi_{it}$ and $\varepsilon_{it}$, $t = 1, \ldots, T_i$, $i = 1, \ldots, N$. Let $f(\cdot)$ denote the probability density function (PDF) for continuous variables and the probability mass function (PMF) for discrete variables. The joint posterior distribution of all unknown parameters and latent variables is proportional to the joint full probability of the unknown parameters, latent variables and the data:

$$
\begin{aligned}
f(\boldsymbol{n}, \boldsymbol{m}, &\boldsymbol{\delta}, \boldsymbol{\lambda}, \boldsymbol{\varepsilon}, \boldsymbol{\xi}, p_0, \boldsymbol{\theta}, \phi, \mu, \sigma^2, \rho, \alpha, \beta, \gamma | \boldsymbol{y}, \boldsymbol{x}, \tilde{\boldsymbol{n}}, \tilde{\boldsymbol{m}}) \\
&\propto f(\boldsymbol{y}, \boldsymbol{n}, \tilde{\boldsymbol{n}}, \boldsymbol{m}, \tilde{\boldsymbol{m}}, \boldsymbol{\delta}, \boldsymbol{\lambda}, \boldsymbol{\varepsilon}, \boldsymbol{\xi}, p_0, \boldsymbol{\theta}, \phi, \mu, \sigma^2, \rho, \alpha, \beta, \gamma | \boldsymbol{x}) \\
&= f(\boldsymbol{y}|\boldsymbol{n}, \boldsymbol{m}, p_0, \boldsymbol{\theta}, \boldsymbol{x}) \times f(\tilde{\boldsymbol{m}}|\tilde{\boldsymbol{n}}, \boldsymbol{\xi}) \times f(\tilde{\boldsymbol{n}}|\boldsymbol{\delta}) \times f(\boldsymbol{m}|\boldsymbol{n}, \boldsymbol{\xi}) \\
&\quad \times f(\boldsymbol{n}|\boldsymbol{\lambda}) \times f(\boldsymbol{\delta}|\boldsymbol{\lambda}, \phi) \times f(\boldsymbol{\lambda}|\mu, \sigma^2, \rho) \times f(\boldsymbol{\varepsilon}|\gamma) \\
&\quad \times f(\mu) \times f(\sigma^2) \times f(\rho) \times f(\phi) \times f(\alpha) \times f(\beta) \\
&\quad \times f(\gamma) \times f(p_0) \times f(\boldsymbol{\theta}),
\end{aligned}
$$
(10)

where $\boldsymbol{\xi}$ exists as a function of $\boldsymbol{\varepsilon}$ given in (8), and known hyper-parameters are suppressed.

To illustrate the MCMC algorithm used to obtain the joint posterior distribution of all parameters, we use VAX004 as an example and give the technical details in the appendix. In summary, we use the following strategies:

- $n$, $m$, $\delta$, $\mu$ and $\sigma^2$ are sampled directly from their full conditional distributions.
- For $\lambda$, $\xi$ and $\varepsilon$, the full conditional distribution is a product of several regular density functions, and we use Metropolized independence sampling with each density sequentially serving as the proposal distribution.
- The random-walk style Metropolis–Hastings algorithm is used for sampling all other parameters.

**4. Application.** In the following, we report the posterior medians followed by the 95% credible sets (CS) for parameters in the Bayesian model, and make comparisons with point estimates followed by the 95% confidence intervals (CI) from the literature when appropriate.

4.1. *VAX004: HIV transmission by sexual contacts.* At each semiannual follow-up visit in trial VAX004, subjects were asked to classify the sexual contacts by the infection status of their partners, that is, positive, negative or unknown, based on their knowledge. HIV prevalence among partners reported as HIV-negative may be less than that among partners reported as HIV-positive. However, an exploratory analysis using a simple likelihood method showed that the probability of infection per contact was not different across the three types of partner infection status reported by the study participants. Hence, we assume a common prevalence $\pi$ of infection among all partners and estimate it by 0.06, the proportion of reported contacts with positive partners among all contacts in the study population. In addition to the analysis for the overall study population, we performed a stratified analysis by classifying the study population into three subgroups corresponding to low, medium and high baseline (month 0) risk levels. We allow the transmission probability and vaccine effect to vary across, but assume that other parameters are not affected by, risk levels. The baseline risk levels are determined by a behavioral risk score ranging from 0 to 7, with 0 as low, 1–3 as medium, and 4–7 as high. The behavioral score is derived from nine baseline risk factors that are highly predictive of HIV infection [Gurwith et al. (2005)].

Table 2 gives the results regarding transmission probabilities and VEs for VAX004. The vaccine did not show a significant effect, reducing the risk of infection per infectious contact by about 7% for the overall study population which is not statistically different from 0. Neither did the low-risk and medium-risk subgroups show any significant vaccine effect. However, we do observe a significant VE of 0.56 (95% CS:0.22, 0.75) in the high-risk subgroup, as the associated 95% CS excludes 0. The pattern that higher baseline risk tends to be associated with higher vaccine efficacy was also identified in Gurwith et al. (2005) via a Cox proportional hazards model for grouped times, where they reported an estimate of 0.06 (95% CS:−0.17, 0.24) for VE per six-month interval for the overall study population and 0.43 (95% CS:0.04, 0.66) for the high-risk subgroup, fairly close to our estimates.

TABLE 2
*VAX004: Summary of the posterior distributions of the transmission probability and the vaccine efficacy per infectious sexual contact for the overall study population and by baseline risk level, compared to the standard analysis*

| Risk level | Total[b] | Infected | $p$ Median | 95% C.S. | VE (Bayesian) Median | 95% C.S. | VE (Cox[a]) Estimate | 95% C.I. |
|---|---|---|---|---|---|---|---|---|
| Overall | 8772 | 368 | 0.0056 | 0.0044, 0.0071 | 0.069 | −0.15, 0.26 | 0.06 | −0.17, 0.24 |
| Low | 3605 | 57 | 0.0020 | 0.0010, 0.0036 | −0.23 | −1.48, 0.35 | −0.48 | −1.93, 0.26 |
| Middle | 4546 | 229 | 0.0054 | 0.0041, 0.0071 | 0.02 | −0.28, 0.25 | 0.03 | −0.25, 0.25 |
| High | 621 | 82 | 0.020 | 0.013, 0.030 | 0.56 | 0.22, 0.75 | 0.43 | 0.04, 0.66 |

[a]Results based on Cox proportional hazards model in Gurwith et al. (2005).
[b]Total number of six-month intervals.

TABLE 3
*VAX004: Summary of the posterior distributions of other parameters for the overall study population*

| Posterior Quantiles | $OR_{con}$ | $\phi$ | $\mu$ | $\sigma^2$ | $\rho$ | $\alpha$ | $\beta$ | $\gamma$ |
|---|---|---|---|---|---|---|---|---|
| Median | 1.44 | 1.66 | −2.54 | 1.95 | 0.92 | 0.30 | 0.29 | 0.65 |
| 2.5% | 1.06 | 1.61 | −2.58 | 1.87 | 0.91 | 0.29 | 0.28 | 0.64 |
| 97.5% | 1.94 | 1.71 | −2.50 | 2.04 | 0.92 | 0.31 | 0.30 | 0.67 |

The baseline transmission probability per infectious sexual contact for the overall study population is 0.0056 (95% CS:0.0044, 0.0071), suggesting that 1000 sexual contacts with HIV-positive partners produce about six infections on average, without intervention of vaccine or condoms. This probability increases across risk levels, with the value for the high risk level 10 times that for the low risk level. A possible reason for the increase in transmission probability across risk levels is that subjects in higher risk levels might more likely under-report the number of contacts.

Results for all other parameters are presented in Table 3. Surprisingly, the reported use of condoms did not seem to be protective with $OR_{con}$ estimated as 1.44 (95% CS:1.06, 1.94), suggesting that it increased the odds of transmission by about 44%. A possible explanation is that the reporting of condom use might be correlated with certain types of sexual behavior. A more specific speculation is that subjects in monogamy tended to use condoms much less frequently and yet had lower risk of infection as compared to those with multiple partners. We included an indicator for monogamy (on average < 2 partners over the study period), but the estimate of OR$_{con}$ did not change much (results not shown).

High within-subject correlation is found among the contact rates and proportions of condom use, with $\rho$ and $\gamma$ estimated as 0.92 (95% CS:0.91, 0.92) and 0.65 (95% CS:0.64, 0.67) respectively. These correlation parameters indicate the magnitude of, but do not directly measure, the correlation coefficients among $\boldsymbol{\lambda}_i$ and among $\boldsymbol{\xi}_i$. Based on posterior medians of $\mu$, $\sigma^2$, $\alpha$ and $\beta$, we found that the mean contact rate in this cohort is 0.21 (95% CS:0.20, 0.22) times per day, and the mean proportion of condom use is 0.51 (95% CS:0.50, 0.52).

If a marginal gamma distribution is assumed for $\boldsymbol{\lambda}_i$, we use the same copula technique used for $\boldsymbol{\xi}_i$ to introduce within-subject correlation. Changing the distribution of the contact rate from log-normal to gamma does not affect the estimates appreciably except for a slight increase in $\phi$ and decrease in $\rho$. We compare predicted population-level means and variances of the reported number of contacts yielded by the two distributions to the observed values, shown in Figure 2(a)–(c). While the gamma distribution gives a predicted mean closer to the observed mean, the log-normal distribution gives a more realistic standard deviation. The heavier tail of the log-normal distribution can better catch extreme reported values. We choose not to ignore the extreme reported values, and therefore, all above results for VAX004 are based on the log-normal distribution for the contact rate.

While we believe that our prior assumptions over most parameters are noninformative or toward-null, we performed a brief sensitivity analysis by changing the prior distribution of $p_0$. We impose a strong beta prior with mean 0.0073 and standard deviation 0.001, instead of Uniform$(0.0001, 0.1)$, on $p_0$. The posterior estimates increase to 0.0063 $(0.0052, 0.0075)$ for $p_0$ and 0.12 $(-0.08, 0.29)$ for VE, and decrease to 1.28 $(0.99, 1.68)$ for OR$_{con}$, all changes being mild. A higher prior mean of $p_0$ will cause more substantial changes in the same directions.

4.2. *VAX003: HIV transmission among IDUs using shared needles.* In the Bayesian probability structure for trial VAX003, the over-dispersion structure and the related parameters, $\phi$ and $\delta_{it}$, are dropped, that is, we assume $\tilde{n}_{it} \sim \text{Poisson}(\lambda_{it})$. The reason is that there is not sufficient information about over-dispersion with only four categories for the contact rate. We stratify the shape and scale parameters by incarceration injection history ($u_i$) for both injection rate ($\lambda_{it}$) and the proportion of needle sharing ($\xi_{it}$), an attempt to control for confounding factors when we evaluate the effect of incarceration injection history on the transmission probability. The prevalence of HIV among IDUs in Bangkok was around 30% [Kitayaporn et al. (1998)]. It was estimated that the relative prevalence between subtypes E and B was growing at a decreasing rate between 1998 and 2000, and reached 70%:30% in 2000 [Kitayaporn et al. (1998), Hudgens et al. (2002)]. Based on this information, the average relative prevalence most likely is between 0.7:0.3 to 0.8:0.2. We use $\pi^{(e)} = 0.75 \times 0.3 = 0.225$ and $\pi^{(b)} = 0.075$ for analyses stratified by subtype.

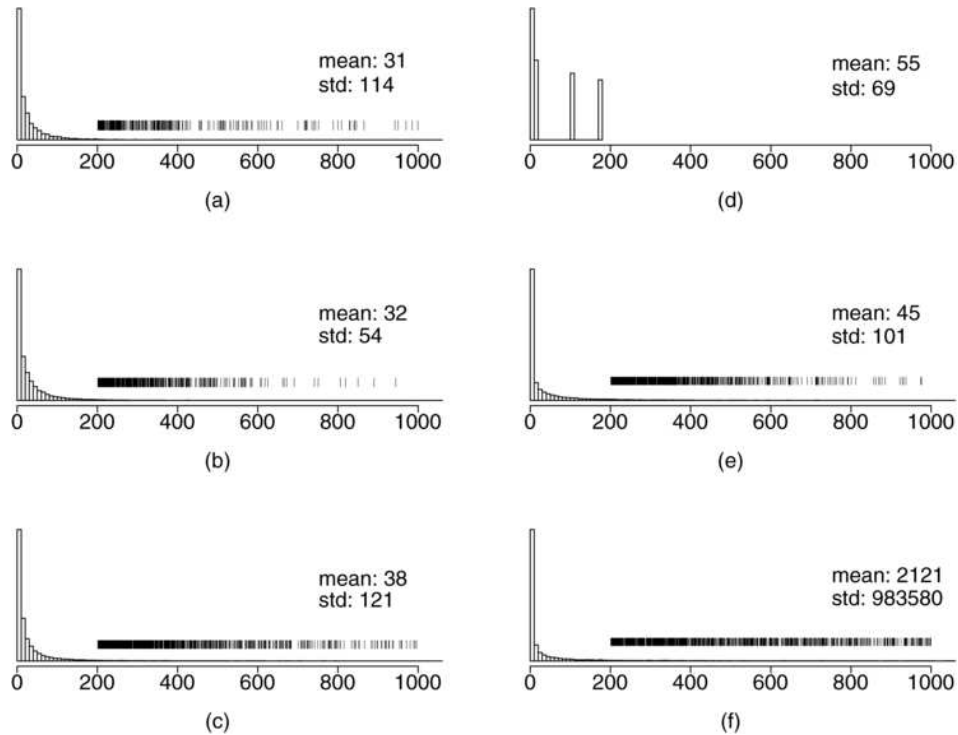

Fig. 2. (a) *Reported number of sexual contacts in VAX004. Values larger than 1000 are truncated. The vertical line segments indicate the location of values between 200 and 1000.* (b) *Predicted number of sexual contacts in VAX004, assuming gamma distribution for contact rate.* (c) *Predicted number of sexual contacts in VAX004, assuming log-normal distribution for contact rate.* (d) *Reported number of injections in VAX003.* (e) *Predicted number of injections in VAX003, assuming gamma distribution for injection rate.* (f) *Predicted number of injections in VAX003, assuming log-normal distribution for injection rate.*

We performed additional analyses stratified by two baseline behavioral risk levels defined in Pitisuttithum et al. (2006). A subject (and all his six-month intervals) is classified into the high baseline risk level if 2 or more of the following risk factors were present at visit 0: use of injection drugs regularly, use of injection drugs daily or weekly, use of injection drugs with shared needles, history of incarceration during the past 6 months, partner was an IDU, or shared needles with partner. Otherwise, the subject is classified into the low risk level.

TABLE 4
*VAX003: Summary of the posterior distributions of the transmission probability and the vaccine efficacy per infectious needle-sharing act for the overall study population and by baseline risk level and HIV subtype, compared to the standard analysis*

| Risk level | Sub-type | Total[b] | Infect-ed[c] | $p$ | | VE (Bayesian) | | VE (Cox[a]) | |
|---|---|---|---|---|---|---|---|---|---|
| | | | | Median | 95% C.S. | Median | 95% C.S. | Estimate | 95% C.I. |
| Overall | | 13797 | 206 | 0.026 | 0.021, 0.031 | −0.08 | −0.43, 0.20 | 0.001 | −0.31, 0.24 |
| | E | | 160 | 0.028 | 0.022, 0.034 | −0.12 | −0.52, 0.17 | −0.014 | −0.38, 0.25 |
| | B | | 32 | 0.019 | 0.012, 0.029 | 0.18 | −0.57, 0.60 | | |
| | E/B | | | 1.45 | 0.91, 2.39 | | | | |
| Low | | 6622 | 80 | 0.033 | 0.024, 0.045 | 0.06 | −0.49, 0.41 | | |
| | E | | 55 | 0.034 | 0.022, 0.048 | 0.04 | −0.66, 0.42 | | |
| | B | | 16 | 0.032 | 0.015, 0.058 | 0.18 | −1.33, 0.67 | | |
| | E/B | | | 1.06 | 0.51, 2.54 | | | | |
| High | | 7175 | 126 | 0.023 | 0.017, 0.029 | −0.10 | −0.60, 0.23 | | |
| | E | | 105 | 0.025 | 0.019, 0.032 | −0.21 | −0.77, 0.19 | | |
| | B | | 16 | 0.015 | 0.008, 0.026 | 0.34 | −0.63, 0.77 | | |
| | E/B | | | 1.68 | 0.92, 3.31 | | | | |

[a]Results based on Cox proportional hazards model in Pitisuttithum et al. (2006).
[b]Total number of six-month intervals.
[c]Intervals for 5 subjects infected by visit 0 (E:4, B:1) are excluded. The 14 untypeable infections are not shown.

The results for transmission probabilities and vaccine efficacies are presented in Table 4. None of the VE estimates are significantly different from 0. We estimate the VE per infectious needle-sharing act as −0.08 (95% CS:−0.43, 0.20) for overall transmission and as −0.12 (95% CS:−0.52, 0.17) for subtype E. Although subtype B tends to have a better VE than subtype E, the difference is not significant. Pitisuttithum et al. (2006) reported similar VE estimates, 0.001 (95% CI:−0.31, 0.24) for the overall IDU cohort and −0.014 (95% CI:−0.38, 0.25) for subtype E, based on a Cox proportional hazards model for grouped times.

The baseline transmission probability per injection using a needle shared with an HIV-positive IDU is 0.026 (95% CS:0.021, 0.031), suggesting that, out of 100 such injections, 2.6 on average will transmit the virus. The subtype-specific baseline transmission probabilities are estimated as 0.028 (95% CS:0.022, 0.034) for $p_0^{(e)}$ and 0.019 (95% CS:0.012, 0.029) for $\hat{p}_0^{(b)}$, higher than the 0.016 (95% CI:0.012, 0.02) and 0.0063 (95% CI:0.0041, 0.0092) estimated in Hudgens et al. (2002) based on a likelihood method. It is interesting that the transmission probability per injection is somewhat higher for the low versus high baseline risk, opposite to the direction observed in VAX004. The ratio of $p_0^{(e)}$ to $p_0^{(b)}$, with a posterior median of 1.45 (95% CS:0.91, 2.39), is only marginally different from 1, lower than the 2.48 (95% CI:1.63, 3.88) reported in Hudgens et al. (2002).

TABLE 5
*VAX003: Summary of the posterior distributions of other parameters for the overall study population*

| Posterior quantiles | $OR_{inc}$ | $\rho$ | $\gamma$ | With incarceration injection history | | | | Without incarceration injection history | | | |
|---|---|---|---|---|---|---|---|---|---|---|---|
| | | | | $\alpha_\lambda{}^a$ | $\beta_\lambda{}^b$ | $\alpha$ | $\beta$ | $\alpha_\lambda{}^a$ | $\beta_\lambda{}^b$ | $\alpha$ | $\beta$ |
| Median | 0.47 | 0.50 | 0.47 | 0.24 | 1.87 | 0.23 | 1.36 | 0.20 | 1.25 | 0.23 | 5.28 |
| 2.5% | 0.30 | 0.48 | 0.44 | 0.21 | 1.60 | 0.20 | 1.12 | 0.19 | 1.18 | 0.22 | 4.85 |
| 97.5% | 0.72 | 0.52 | 0.51 | 0.27 | 2.24 | 0.26 | 1.66 | 0.21 | 1.32 | 0.25 | 5.75 |

[a]Shape of the gamma distribution for contact rate.

[b]Scale of the gamma distribution for contact rate.

Table 5 summarizes estimates for all other parameters. The odds ratio for incarceration injection is estimated as 0.47 (95% CS:0.30, 0.72). Hudgens et al. (2002) reported a much higher value, 4.47 (95% CI:2.63, 7.19), where a time-varying prevalence ratio with an average about 0.55:0.45 between subtypes E and B and a common proportion of 4% for needle sharing across the whole population were assumed. Among subjects with incarceration injection history, the mean injection rate is 0.45 (95% CS:0.37, 0.54) times per day and 14% (95% CS:12%, 17%) involved shared needles. In contrast, among those without incarceration history, the mean injection rate is 0.25 (95% CS:0.24, 0.27) times per day and 4.2% (95% CS:4.0%, 4.5%) involved shared needles. The assumption of a common proportion of needle-sharing in Hudgens et al. (2002) lowers the injection frequency and proportion of needle-sharing down to the overall level, and consequently increases the adjusted transmission probability for subjects with incarceration history. In addition, the incarceration injection indicator is defined for each interval in Hudgens et al. (2002), whereas we define it for each individual. Posterior estimates of $\rho$, 0.50 (95% CS:0.48, 0.52), and $\gamma$, 0.47 (95% CI:0.44, 0.51), suggest substantial within-subject correlation, though not as high as those in VAX004.

Similar to VAX004, log-normal and gamma distributions for the injection rate lead to similar results, with a slight difference in $\rho$. In Figure 2(d)–(f), we see that the heavy tail of the log-normal distribution yields extremely large predicted moments for the reported number of injections and thus makes it less competitive than the gamma distribution for modeling injection rates reported in a few categories. Consequently, all results presented for VAX003 are based on the gamma distribution for injection rate.

We performed sensitivity analyses by changing the relative prevalence $\pi^{(e)}:\pi^{(b)}$ to 0.7:0.3 and 0.8:0.2. As expected, the transmission probability tends to decrease for subtype E but to increase for subtype B, as the relative prevalence of subtype E increases. For each subtype and risk level, the VE estimate changes in the direction opposite to that of the corresponding transmission probability, but none of the VE estimates differ significantly from 0. The magnitude of all these changes are relatively small, especially for subtype E. The estimated transmission probability ratio of subtype E to subtype B decreases as the relative prevalence of subtype E increases. Particularly, subtype E becomes statistically more infectious than subtype B with an estimate of 1.88 (95% CS: 1.18, 3.21) for $p_0^{(e)}/p_0^{(b)}$, if the prevalence of subtype E is as low as 70% among the IDUs.

**5. Discussion.** We established a Bayesian hierarchical model for analyzing clinical studies of infectious disease with transmission and exposure data observed over discrete time intervals. This model provides assessment of the transmission probability and vaccine efficacy conditioning on an infectious contact, whereas standard methods of analyzing vaccine trials do not. Assuming conditional independence between observed and true but unobserved quantities, this model provides an approach to adjustment for the measurement error in some key risk factors. We used the method to reanalyze two HIV-1 vaccine trials on populations who are at high risk of HIV-transmission via sexual contacts or sharing needles for drug injection. The proposed method could be applied to studies of other vaccines, such as human papilloma virus vaccines, where contact information is collected.

We obtained estimates of vaccine efficacy similar to the primary study results, especially for VAX004, confirming the findings of no protective efficacy. Two factors may contribute to this similarity in VE estimates. First, the measurement error might be relatively small for the majority of the study population. Second, our model assumes unbiasness, that is, $\mathrm{E}(\tilde{n}_{it}|\lambda_{it}) = \mathrm{E}(n_{it}|\lambda_{it})$ and $\mathrm{E}(\tilde{m}_{it}|\lambda_{it},\xi_{it}) = \mathrm{E}(m_{it}|\lambda_{it},\xi_{it})$. However, if the bias trend is similar in both treatment groups, even a model with bias correction will likely yield a similar VE estimate as well. Despite the similarity, our hierarchical model provides joint inference on not only the transmission probability and VE but also the population-level behavioral characteristics such as the contact rate and proportion of condom use (needle-sharing).

We have assumed an exchangeable structure for within-subject correlation among contact (injection) rates and proportions of condom use (needle-sharing), using the copula method. A more sophisticated structure may be considered given sufficient data. Within-subject sample correlation coefficients among the logarithm of reported contact rates, $\{\log(\tilde{n}_{it}/l_{it}) : t = 1,\ldots,T_i\}$, and among reported proportions of condom use, $\{\tilde{m}_{it}/\tilde{n}_{it} : t = 1,\ldots,T_i\}$, in VAX004 do indicate that correlation wanes away as two intervals are further apart, but the variation range is relatively small, 0.3–0.5 for the former and 0.45–0.67 for the latter. Therefore, an exchangeable structure is a reasonable assumption, albeit an autoregressive structure such as the ARMA$(p,q)$ model [Chib and Greenberg (1994)] may be more realistic. The range of 0.3–0.5 for $\{\log(\tilde{n}_{it}/l_{it}) : t = 1, \ldots, T_i\}$ may seem contradictory to the Bayesian estimate of $\rho$ around 0.9. A plausible explanation is that the addition of $\delta_{it}$ to reflect the over-dispersion may attenuate the true correlation among the elements of $\boldsymbol{\lambda}_i$, as the elements of $\boldsymbol{\delta}_i$ are independent given $\boldsymbol{\lambda}_i$. Consequently, a high correlation among the elements of $\boldsymbol{\lambda}_i$ is needed to yield a moderate marginal correlation among the elements of $\boldsymbol{\delta}_i$. In fact, the parameter estimates, especially for transmission probabilities and VEs, do not change much if we assume intervals within the same subject are independent. A possible reason is that only the overall magnitude of $\boldsymbol{n}_i$ and $\boldsymbol{m}_i$ matter in the estimation of $p_0$ and the VE, and the magnitude mainly depends on the observed $\tilde{\boldsymbol{n}}_i$ and $\tilde{\boldsymbol{m}}_i$ and is much less affected by the correlation. However, we do see that correlation adjustment changes the shape and scale of the distributions of the contact rate $\lambda_{it}$ and the proportion $\xi_{it}$ in a more noticeable way. For example, without incarceration injection history, the estimates for the shape parameter $\beta$ for the proportion of needle-sharing in VAX003 change from 5.28 (95% CS:4.85, 5.75) to 6.3 (95% CS:5.92, 6.69) when within-subject independence is assumed.

To adjust for error in self-reported contact information, we assumed a Poisson process for the true number of contacts and an over-dispersed Poisson process for the reported one, and that the two processes are conditionally independent given the underlying contact rate. Ideally, validation data would be available so that the measurement error could be modeled parametrically or without parametric assumptions as in Golm, Halloran and Longini (1999). The collection of validation data would be useful in future vaccine trials. In this Bayesian framework, a more general bivariate distribution could be modeled between $n_{it}$ and $\tilde{n}_{it}$ given $\lambda_{it}$ or between $\lambda_{it}$ and a latent rate $\tilde{\lambda}_{it}$ that determines the distribution of $\tilde{n}_{it}$, had validation data been available on contact frequency. Another form of additional data, replication of $\tilde{n}_{it}$ and $\tilde{m}_{it}$ in all or some of the intervals, can also improve model precision [Carroll, Ruppert and Stefanski (1995)], but the assumption of unbiasness of $\tilde{n}_{it}$ for the true $n_{it}$ has to be retained. A possible parametric utilization of replication data in our model is to allow for within-interval correlation.

Other than log-normal and gamma distributions, a more flexible option for modeling the contact rate may be mixture prior densities [Richardson et al. (2002)]. It is likely that the true number of contacts also comes from an over-dispersed Poisson process, but whether such a model is identifiable needs further investigation. When the number of contacts is given as $K$ categories and $K$ is small, for example, in trial VAX003, the Poisson and over-dispersed Poisson structure may not be realistic. In that case, a more flexible probability structure is to assume that $n_{it}$ and $\tilde{n}_{it}$ independently follow a discrete distribution indexed by $\boldsymbol{p}_{it} = (p_{it1}, \ldots, p_{itK})^\tau$, where $p_{itk}$ is the probability of falling in the $k$th category for interval $(i,t)$, and $\boldsymbol{p}_{it} \sim$ Dirichlet($\boldsymbol{\alpha}$) for some random or known vector $\boldsymbol{\alpha}$.

The model is sensitive to the contact-related information when such information is limited. For instance, when the value assigned to the "None" category of the reported proportion of needle-sharing was increased from 0.5% to 5% or higher, we were unable to obtain convergence, likely due to the lack of curvature supporting the estimation of a beta density. We emphasize for future studies that, in terms of contact frequency, numbers are more informative than categories, and more categories are preferred to fewer. Another factor to which the analyses are sensitive is the prevalence of infections among partners. While it is impossible to obtain the infection status of all partners, a validation set of partners randomly selected for verification of infection would help improve the inference. To alleviate under- or over-reporting of contact frequency, it is also important to ensure that study participants understand the definition of a contact, especially when the study involves multiple contact types. Extremely high frequencies, for example, the numbers of sexual contacts that were reported as over thousands per six-month interval by several participants in VAX004, may indicate misunderstanding of the definition, and should be verified with the participants during the follow-up visits. The underlying mechanism of measurement error in contact-related factors in real studies may never be known, and the best way to improve the VE estimation is to reduce the error at the data collection step.

## APPENDIX: MCMC METHODS AND RELATED SAMPLING ISSUES

**MCMC sampling schemes.** We use $f_{dist}(\cdot|\cdot)$ to denote the PDF for continuous variables or the PMF for discrete variables, and $F_{dist}(\cdot|\cdot)$ to denote the CDF of a random variable given parameters. The subscript "$dist$" could be "$Bin$" for binomial, "$Pois$" for Poisson, "$Beta$" for beta, "$G$" for gamma, "$IG$" for inverse gamma, "$N$" for normal and "$LN$" for log-normal distributions. Whether the distribution is univariate or multivariate is determined by the parameter input.

**Sampling $n_{it}$.** Define $q_{i1} = 1 - p(v_i, 1)$ as the probability of escaping infection from a contact protected by condom use, and similarly, define $q_{i0} = 1 - p(v_i, 0)$ for an un-protected contact. The conditional probability of $n_{it}$ is

20    Y. YANG, P. GILBERT, I. M. LONGINI, JR. AND M. E. HALLORANgiven by

$$\Pr(n_{it} = n|\cdot) = \begin{cases} \dfrac{(\lambda_{it}l_{it}(1-\xi_{it})q_{i0})^{n-m_{it}}\exp\{-\lambda_{it}l_{it}(1-\xi_{it})q_{i0}\}}{(n-m_{it})!}, & y_{it} = 0, \\ \dfrac{(\lambda_{it}l_{it}(1-\xi_{it}))^{n-m_{it}}\exp\{-\lambda_{it}l_{it}(1-\xi_{it})\}}{(n-m_{it})!} \\ \quad \times [1 - q_{i1}^{m_{it}}q_{i0}^{n-m_{it}}]/C_{it}, & y_{it} = 1, \end{cases}$$

where $C_{it} = 1 - q_{i1}{}^{m_{it}}\exp\{\lambda_{it}l_{it}(1-\xi_{it})(q_{i0}-1)\}$. When $y_{it} = 0$, we sample $n_{it} - m_{it}$ directly from $\text{Poisson}(\lambda_{it}l_{it}(1-\xi_{it})q_{i0})$. When $y_{it} = 1$, note that the conditional CDF of $n_{it}$ is

(11)
$$\begin{aligned}\Pr(n_{it} \leq n|\cdot, y_{it} = 1) &= [\exp\{\lambda_{it}l_{it}(1-\xi_{it})\}F_{\text{Pois}}(n|\lambda_{it}l_{it}(1-\xi_{it})) \\ &\quad - q_{i1}{}^{m_{it}}\exp\{\lambda_{it}l_{it}(1-\xi_{it})q_0\}F_{\text{Pois}}(n|\lambda_{it}l_{it}(1-\xi_{it})q_{i0})] \\ &\quad \times [\exp\{\lambda_{it}l_{it}(1-\xi_{it})\} - q_{i1}{}^{m_{it}}\exp\{\lambda_{it}l_{it}(1-\xi_{it})q_{i0}\}]^{-1}.\end{aligned}$$

As the CDF is a nondecreasing function, we use direct sampling in combination with binary searching. For example, to sample $n_{it}$, we generate a value $z$ from $\text{Uniform}(0,1)$; then, the smallest $n$ satisfying $\Pr(n_{it} \leq n|\cdot, y_{it} = 1) \geq z$ is the sampled value of $n_{it}$ and can be found using binary searching or other advanced searching methods.

**Sampling $m_{it}$.** The conditional probability of $m_{it}$ is

$$\Pr(m_{it} = m|\cdot) = \begin{cases} \binom{n_{it}}{m}\dfrac{\xi_{it}q_{i1}}{\xi_{it}q_{i1} + (1-\xi_{it})q_{i0}}^m \\ \quad \times \left(1 - \dfrac{\xi_{it}q_{i1}}{\xi_{it}q_{i1} + (1-\xi_{it})q_{i0}}\right)^{n_{it}-m}, & y_{it} = 0, \\ \binom{n_{it}}{m}\xi_{it}^m(1-\xi_{it})^{n_{it}-m} \times [1 - q_{i1}^m q_{i0}^{n_{it}-m}]/D_{it}, & y_{it} = 1, \end{cases}$$

where $D_{it} = 1 - [\xi_{it}q_{i1} + (1-\xi_{it})q_{i0}]^{n_{it}}$. When $y_{it} = 0$, we sample $m_{it}$ directly from $\text{Binomial}(n_{it}, \frac{\xi_{it}q_{i1}}{\xi q_{i1}+(1-\xi_{it})q_{i0}})$. When $y_{it} = 1$, we have

(12)
$$\Pr(m_{it} \leq m|\cdot, y_{it} = 1) = \frac{F_{Bin}(m|n_{it}, \xi_{it}) - (\xi q_{i1} + (1-\xi_{it})q_{i0})^{n_{it}}F_{Bin}(m|n_{it}, \mathcal{P})}{1 - (\xi_{it}q_{i1} + (1-\xi_{it})q_{i0})^{n_{it}}},$$

where $\mathcal{P} = \frac{\xi_{it}q_{i1}}{\xi_{it}q_{i1}+(1-\xi_{it})q_{i0}}$. We use the same technique in sampling $n_{it}$, that is, direct sampling in combination with binary searching.

**Sampling $\lambda_{it}$.** Define $\boldsymbol{\mu}_i = \mu \mathbf{1}_{T_i \times 1}$ and $\boldsymbol{\Sigma}_i = \sigma^2(\rho \boldsymbol{J}_{T_i \times T_i} + (1-\rho)\boldsymbol{I}_{T_i \times T_i}))$. The likelihood part concerning the contact rate vector $\boldsymbol{\lambda}_i$ is given by

$$L_i(\boldsymbol{\lambda}_i|\cdot) \propto f_{LN}(\boldsymbol{\lambda}_i|\boldsymbol{\mu}_i, \boldsymbol{\Sigma}_i)\left(\prod_{t=1}^{T_i} f_G(\lambda_{it}|n_{it}+1, l_{it}^{-1})\right)$$

$$\times \left(\prod_{t=1}^{T_i} f_{IG}\left(\lambda_{it}\Big|\phi, \frac{l_{it}}{\phi\delta_{it}}\right)\right)\left(\prod_{t=1}^{T_i} \lambda_{it}\right).$$

To sample $\boldsymbol{\lambda}_i$, we take the following steps:

- First sample $\boldsymbol{\lambda}_i^\star$ from Log-Normal$(\boldsymbol{\mu}_i, \boldsymbol{\Sigma}_i)$, and accept it with the probability

$$\min\left(1, \frac{\prod_{t=1}^{T_i}\{f_G(\lambda_{it}^\star|n_{it}+1, l_{it}^{-1})f_{IG}(\lambda_{it}^\star|\phi, l_{it}/(\phi\delta_{it}))\lambda_{it}^\star\}}{\prod_{t=1}^{T_i}\{f_G(\lambda_{it}|n_{it}+1, l_{it}^{-1})f_{IG}(\lambda_{it}|\phi, l_{it}/(\phi\delta_{it}))\lambda_{it}\}}\right).$$

Update $\boldsymbol{\lambda}_i$ with $\boldsymbol{\lambda}_i^\star$ if the new sample is accepted;
- Sample a new $\boldsymbol{\lambda}_i^\star$ from $\prod_{t=1}^{T_i}(f_G(\lambda_{it}^\star|n_{it}+1, l_{it}^{-1})$, and accept it with the probability

$$\min\left(1, \frac{f_{LN}(\boldsymbol{\lambda}_i^\star|\boldsymbol{\mu}_i, \boldsymbol{\Sigma}_i)\prod_{t=1}^{T_i}\{\lambda_{it}^\star f_{IG}(\lambda_{it}^\star|\phi, l_{it}/(\phi\delta_{it}))\}}{f_{LN}(\boldsymbol{\lambda}_i|\boldsymbol{\mu}_i, \boldsymbol{\Sigma}_i)\prod_{t=1}^{T_i}\{\lambda_{it} f_{IG}(\lambda_{it}|\phi, l_{it}/(\phi\delta_{it}))\}}\right).$$

Update $\boldsymbol{\lambda}_i$ with $\boldsymbol{\lambda}_i^\star$ if the new sample is accepted;
- Sample a new $\boldsymbol{\lambda}_i^\star$ from $\prod_{t=1}^{T_i} f_{IG}(\lambda_{it}|\phi, \frac{l_{it}}{\phi\delta_{it}})$ and accept it with the probability

$$\min\left(1, \frac{f_{LN}(\boldsymbol{\lambda}_i^\star|\boldsymbol{\mu}_i, \boldsymbol{\Sigma}_i)\prod_{t=1}^{T_i}\{\lambda_{it}^\star f_G(\lambda_{it}^\star|n_{it}+1, l_{it}^{-1})\}}{f_{LN}(\boldsymbol{\lambda}_i|\boldsymbol{\mu}_i, \boldsymbol{\Sigma}_i)\prod_{t=1}^{T_i}\{\lambda_{it} f_G(\lambda_{it}|n_{it}+1, l_{it}^{-1})\}}\right).$$

This cross-sampling procedure is a generalization of the Metropolized independence sampling algorithm [Chib and Greenberg (1995)]. Liu (1996) showed that Metropolized independence sampling is superior to rejection sampling with respect to asymptotic efficiency and ease of computation, given that the proposal density provides a reasonable coverage over the domain of the posterior density. In this case, we have a composite full likelihood $L(x) \propto f(x)g(x)$ in which $f(x)$ and $g(x)$ are both ready for sampling. Using $f(x)$ and $g(x)$ alternately as the proposal density can better cover the reasonable range of $x$ as compared to using either $f(x)$ or $g(x)$ alone as the proposal density.

**Sampling $\boldsymbol{\varepsilon}_i$ and $\boldsymbol{\xi}_i$.** Define $\boldsymbol{\Upsilon}_i = \gamma \boldsymbol{J}_{T_i \times T_i} + (1-\gamma)\boldsymbol{I}_{T_i \times T_i}$. The likelihood part concerning $\boldsymbol{\varepsilon}_i$ is given by

$$(13) \quad L_i(\boldsymbol{\varepsilon}_i|\cdot) \propto f_N(\boldsymbol{\varepsilon}_i|\mathbf{0}, \boldsymbol{\Upsilon}_i) \times \prod_{t=1}^{T_i} \xi_{it}^{m_{it}+\tilde{m}_{it}}(1-\xi_{it})^{n_{it}-m_{it}+\tilde{n}_{it}-\tilde{m}_{it}}.$$

The above likelihood is expressed in terms of $\boldsymbol{\varepsilon}_i$, and $\boldsymbol{\xi}_i$ exists through $\xi_{it} = \Psi^{-1}(\Phi(\varepsilon_{it})|\alpha,\beta)$. To express the likelihood in terms of $\boldsymbol{\xi}_i$, (13) becomes

$$L_i(\boldsymbol{\xi}_i|\cdot) \propto \exp\{-\tfrac{1}{2}\boldsymbol{\varepsilon}_i^\tau \boldsymbol{\Upsilon}_i^{-1}\boldsymbol{\varepsilon}_i + \tfrac{1}{2}\boldsymbol{\varepsilon}_i^\tau \boldsymbol{\varepsilon}_i\}$$
$$\times \prod_{t=1}^{T_i} f_{Beta}(\boldsymbol{\xi}_i|\alpha + m_{it} + \tilde{m}_{it}, \beta + n_{it} - m_{it} + \tilde{n}_{it} - \tilde{m}_{it}),\tag{14}$$

where $\boldsymbol{\varepsilon}_i$ exists via $\varepsilon_{it} = \Phi^{-1}(\Psi(\xi_{it}|\alpha,\beta))$. $|\prod_{t=1}^{T_i}\{\Phi^{-1\prime}(\Psi(\xi_{it}))\Psi'(\xi_{it})\}|$ is the Jacobian term, and $\Phi^{-1\prime}(x) = [f_N(\Phi^{-1}(x)|0,1)]^{-1}$.

The sampling of $\boldsymbol{\varepsilon}_i$ and $\boldsymbol{\xi}_i$ proceeds as the following:

- Based on (13), sample $\boldsymbol{\varepsilon}_i^\star$ from Normal($\boldsymbol{0}, \boldsymbol{\Upsilon}_i$), and accept it with the probability

$$\min\left(1, \prod_{t=1}^{T_i} \frac{\xi_{it}^{\star\,m_{it}+\tilde{m}_{it}}(1-\xi_{it}^\star)^{n_{it}-m_{it}+\tilde{n}_{it}-\tilde{m}_{it}}}{\xi_{it}^{m_{it}+\tilde{m}_{it}}(1-\xi_{it})^{n_{it}-m_{it}+\tilde{n}_{it}-\tilde{m}_{it}}}\right),$$

where $\xi_{it}^\star = \Psi^{-1}(\Phi(\varepsilon_{it}^\star))$. Update $\boldsymbol{\varepsilon}_i$ and $\boldsymbol{\xi}_i$ if the new sample is accepted;
- Based on (14), sample $\boldsymbol{\xi}_i^\star$ from $\prod_{t=1}^{T_i} f_{Beta}(\alpha + m_{it} + \tilde{m}_{it}, \beta + (n_{it} - m_{it}) + (\tilde{n}_{it} - \tilde{m}_{it}))$, and accept it with the probability

$$\min\left(1, \frac{\exp\{(-1/2)\boldsymbol{\varepsilon}_i^{\star\tau}\boldsymbol{\Upsilon}_i^{-1}\boldsymbol{\varepsilon}_i^\star + (1/2)\boldsymbol{\varepsilon}_i^{\star\tau}\boldsymbol{\varepsilon}_i^\star\}}{\exp\{(-1/2)\boldsymbol{\varepsilon}_i^\tau\boldsymbol{\Upsilon}_i^{-1}\boldsymbol{\varepsilon}_i + (1/2)\boldsymbol{\varepsilon}_i^\tau\boldsymbol{\varepsilon}_i\}}\right),$$

where $\varepsilon_{it}^\star = \Phi^{-1}(\Psi(\xi_{it}^\star))$.

**Sampling other parameters.** Let $\log \boldsymbol{\lambda}_i = (\log \lambda_{i1}, \ldots, \log \lambda_{iT_i})^\tau$, $\boldsymbol{\mu}_i = \mu \mathbf{1}_{T_i \times 1}$, and let $\boldsymbol{R}_i = \rho \boldsymbol{J}_{T_i \times T_i} + (1-\rho)\boldsymbol{I}_{T_i \times T_i}$ such that $\boldsymbol{\Sigma}_i = (\sigma^2)\boldsymbol{R}_i$.

The following parameters are sampled directly from their full conditional distributions:

$$\delta_{it}|\cdot \sim \text{Gamma}\left(\tilde{n}_{it} + \phi, \left(1 + \frac{\phi}{\lambda_{it}l_{it}}\right)^{-1}\right),$$

$$\mu|\cdot \sim \text{Normal}\left(\frac{\sum_{i=1}^N \mathbf{1}^\tau \boldsymbol{\Sigma}_i^{-1} \log \boldsymbol{\lambda}_i}{\sum_{i=1}^N \mathbf{1}^\tau \boldsymbol{\Sigma}_i^{-1} \mathbf{1}}, \left(\sum_{i=1}^N \mathbf{1}^\tau \boldsymbol{\Sigma}_i^{-1} \mathbf{1}\right)^{-1}\right),$$

$$\sigma^2|\cdot \sim \text{Inverse Gamma}\left(\tfrac{1}{2}\sum_i T_i, \left[\tfrac{1}{2}\sum_{i=1}^N (\log \boldsymbol{\lambda}_i - \boldsymbol{\mu}_i)^\tau \boldsymbol{R}_i^{-1}(\log \boldsymbol{\lambda}_i - \boldsymbol{\mu}_i)\right]^{-1}\right).$$

A random-walk style Metropolis–Hastings algorithm is used to sample $\rho$, $\phi$, $\alpha$, $\beta$, $\gamma$, $p_0$ and $\boldsymbol{\theta}$, that is, a new value is sampled from a normal density with the current value as its mean. The variance of each proposal normal density is dynamically adapted to reach an acceptance rate of 0.3–0.4. To apply this sampling scheme, appropriate transformation may be necessary so that the domain of the transformed parameter is $(-\infty, \infty)$, for example, a logit transformation for the transmission probability.

**Diagnostics for convergence.** We run three chains simultaneously and use the scale reduction factor to monitor the convergence of the chains. The scale reduction factor is defined as

$$\sqrt{\hat{R}} = \sqrt{\frac{M-1}{M} + \frac{1}{M}\frac{B}{W}},$$

where $M$ is the number of runs, and $B$ and $W$ are the between-sequence and within-sequence variances, respectively. Gelman and Rubin (1992) showed that the factor $\sqrt{\hat{R}}$ will approach 1 as $M \to \infty$, and recommended that the convergence can be considered as reached if $\sqrt{\hat{R}} < 1.2$ for all parameters. We calculate $\sqrt{\hat{R}}$ for each 5000 iterations afterward and the criteria $\sqrt{\hat{R}} < 1.2$ is adopted as the stopping rule.

The results of analyzing the two AIDSVAX trials are based on the last 5000 iterations of three parallel chains. A burn-in period of 5000 runs is enforced after the variances of proposal normal densities are fixed. To reduce the correlation within each successive chain, we loop over the last 5000 runs of the three parallel chains, and at each loop we randomly pick one chain to read in the samples.

Y. Yang  
Program of Biostatistics  
  and Biomathematics  
Division of Public Health Sciences  
Fred Hutchinson Cancer Research Center  
Seattle, Washington 98109  
USA  
E-mail: yang@fhcrc.org

P. Gilbert  
I. M. Longini, Jr.  
M. E. Halloran  
Department of Biostatistics  
School of Public Health  
  and Community Medicine  
University of Washington  
Seattle, Washington 98195  
USA